\documentclass[preprint,5p,twocolumn]{elsarticle}
\usepackage{amssymb}
\usepackage{amsmath}
\usepackage{mathrsfs}
\usepackage{natbib}

\usepackage{epsfig}
\usepackage{graphics}
\usepackage{float}
\usepackage{tabularx}
\usepackage{color}
\expandafter\ifx\csname package@font\endcsname\relax\else
 \expandafter\expandafter
 \expandafter\usepackage
 \expandafter\expandafter
 \expandafter{\csname package@font\endcsname}%
\fi
\newcolumntype{C}[1]{>{\centering\arraybackslash}p{#1}}

\begin{document}
 

\title{Power law Starobinsky model of inflation from no-scale SUGRA}
\author[PRL]{Girish Kumar Chakravarty}
\ead{girish20@prl.res.in}
\author[PRL]{Subhendra Mohanty}
\ead{mohanty@prl.res.in}

\address[PRL]{Theoretical Physics Division, Physical Research Laboratory, Ahmedabad 380009, India.}

\def\be{\begin{equation}}
\def\ee{\end{equation}}
\def\al{\alpha}
\def\ba{\begin{eqnarray}}
\def\ea{\end{eqnarray}}
\def\beas{\begin{eqnarray*}}
\def\eeas{\end{eqnarray*}}


\begin{abstract}
We consider a power law $\frac{1}{M^2}R^{\beta}$ correction to Einstein gravity as a model for inflation. 
The interesting feature of this form of generalization is that small deviations from the Starobinsky limit
$\beta=2$ can change the value of tensor to scalar ratio from $r \sim \mathcal{O}(10^{-3})$ to $r\sim \mathcal{O}(0.1)$.
We find that in order to get large tensor perturbation $r\approx 0.1$ as  indicated  by BKP measurements, we
require the value of $\beta \approx 1.83$ thereby breaking global Weyl symmetry. We show that the general
$R^\beta$ model can be obtained from a SUGRA construction by adding a power law $(\Phi +\bar \Phi)^n$ term to
the minimal no-scale SUGRA K\"ahler potential. We further show that this two parameter power law generalization of the 
Starobinsky model is equivalent to generalized non-minimal curvature coupled models with quantum corrected $\Phi^{4}$- 
potentials i.e. models of the form $\xi \Phi^{a} R^{b} + \lambda \Phi^{4(1+\gamma)}$ and thus the power law Starobinsky 
model is the most economical parametrization of such models.
\end{abstract}


\maketitle

\section{Introduction}
The Starobinsky model of inflation  \cite{starobinsky1,starobinsky2} with an $\frac{1}{M^{2}}R^{2}$ interaction term is of interest
as it requires no extra scalar fields but relies on the scalar degree of the metric tensor to generate the 'inflaton' potential.
The $R^{2}$ Starobinsky model gives rise to a 'plateau potential' of the inflaton when transformed to the Einstein frame.
This model was favored by the Planck constraint on the tensor to scalar ratio which ruled out potentials like $m^{2}\phi^{2}$ and
$\lambda \phi^{4}$. In addition the Starobinsky model could be mapped to the Higgs-inflation models with 
$\xi \phi^{2} R + \lambda \phi^{4}$ theory \cite{Bezrukov}. The characteristic  feature of the Starobinsky 
equivalent models was the prediction that the tensor to scalar ratio was $r \simeq 10^{-3}$.
BICEP2 reported a large value of $r= 0.2^{+0.07}_{-0.05}$ \cite{BICEP2} but the recent joint analysis
by Planck + BICEP2 + Keck Array give only an upper bound of $r_{0.05}<0.12 (95\% CL)$ \cite{BKP:2015,Ade:2015lrj,Planck:2015}.
In an analysis of the genus structure of the B-mode polarisation of Planck + BICEP2 data by Colley et al. put the tensor to scalar ratio
at $r=0.11\pm0.04 (68\% CL)$ \cite{Colley:2014nna}. In the light of the possibility that $r$ can be larger than the
Starobinsky model prediction of $r\sim0.003$, generalisations of the Starobinsky model are of interest.
%
%

We study a general power law $\frac{1}{6 M^{2}} \frac{R^{\beta}}{M_p^{2\beta-2}}$ correction to the Einstein gravity and compute 
the scalar and tensor power spectrum as a function of the two dimensionless parameters $M$ and $\beta$.  It is well known that the 
$\frac{1}{M^{2}}R^{2}$ model is equivalent to the $\xi \phi^{2} R + \lambda \phi^{4}$ Higgs-inflation model as they led to the
same scalar potential in the Einstein frame \cite{faraoni,Kehagias:2013mya}. One can find similar equivalence between
generalized Higgs-inflation models and the power law Starobinsky model whose common feature is violation of the global Weyl symmetry.
A general scalar curvature coupled $\xi \phi^{a} R^{b}$ model was studied in \cite{girish}. The quantum correction on
$\phi^{4}$-potential in Jordan frame was studied in \cite{Joergensen:2014rya,Codello:2014sua,Gao:2014fha} where they have shown the
equivalence of the $\xi \phi^{2} R + \lambda \phi^{4(1+\gamma)}$ model with $\frac{1}{M^{2}}R^{\beta}$ model. 
The generalized Starobinsky model with $R^{p}$ correction has been studied in the ref.
\cite{stelle,Barrow:1988,Barrow:1991hg,Martin:2013tda,Sebastiani:2013eqa,Costa:2014lta,Cai:2014bda}. In general scalar-curvature theories
the scalar plays the role of the inflaton after transforming to Einstein frame whereas in pure curvature theories like $R + \frac{1}{M^{2}}R^{\beta}$
model the longitudinal part of the graviton is the equivalent scalar in the Einstein frame plays the role of inflaton.

The higher order curvature theories arise naturally in theories of supergravity. 
The supergravity embedding of the Higgs-inflation \cite{Bezrukov} does not produce a slow roll
potential in MSSM  but a potential suitable for inflation is obtained in NMSSM \cite{Einhorn:2009bh}.
The potential in NMSSM however has a tachyonic instability in the direction orthogonal to the slow roll
\cite{Ferrara:2010yw}. This instability can be cured by the addition of quartic terms of the fields  in the
K\"ahler potential \cite{Lee:2010hj,Ferrara:2010in}. 

In the context of a supergravity embedding of the Starobinsky model, It was shown by Cecotti\cite{Cecotti:1987sa} that 
quadratic Ricci  curvature terms  can be derived in a supergravity theory by adding two chiral superfields in the minimal supergravity. 
A no-scale SUGRA\cite{Cremmer:1983bf,Ellis:1984bm,Lahanas:1986uc} model with a modulus field and the inflation field with
a minimal Wess-Zumino superpotential gives the same F-term potential in the Einstein frame as the Starobinsky model
\cite{Ellis:2013xoa}. The symmetry principle which can be invoked for the SUGRA generalization
of the Starobinsky model is the spontaneous violation of superconformal symmetry \cite{Kallosh:2013lkr}. The quadratic
curvature can also arise from D-term in a minimal-SUGRA theory with the addition of a vector and chiral
supermultiplets \cite{Cecotti:1987qe}. The Starobinsky model has been derived from the D-term potential of 
a SUGRA model \cite{Buchmuller:2013zfa,Ferrara:2013rsa,Farakos:2013cqa}. Quartic powers of Ricci curvature in 
the bosonic Lagrangian can also be obtained in a SUGRA model by the D-term of higher 
order powers of the field strength superfield \cite{Farakos:2013cqa,Ferrara:2013kca}.

In this paper we give a SUGRA model for the general power law $\frac{1}{M^{2}}R^\beta$ model. We show that adding
a $(\Phi+\bar {\Phi})^n$ term to the minimal no-scale K\"ahler potential and with a Wess-Zumino form
of the superpotential $W(\Phi)$ yields the same potential in the Einstein frame as the generalised 
Starobinsky model. In the limit $n=2$ the Starobinsky limit $\beta=2$ is obtained. We derive the 
relations between the two parameters of the power-law Starobinsky model and the two parameters 
of our SUGRA model. The interesting point about the generalization is that small deviations from 
the Starobinsky limit of $n=\beta=2$ can produce large shifts in the values of $r$. Many SUGRA models
have been constructed which can yield a range of $r$ from $10^{-3}-10^{-1}$ by changing the parameters 
of the K\"ahler potential and the superpotential \cite{Ferrara:2013kca,Kallosh:2010ug,Nakayama:2013jka,Nakayama:2013txa,
Li:2013nfa,Pallis:2013yda,Cecotti:2014ipa,Ferrara:2014ima,Pallis:2014dma,Harigaya:2014qza,
Ellis:2014rxa,Hamaguchi:2014mza,linde1,Ellis:2014gxa,Ellis:2014opa,Ellis:2014dxa,Diamandis:2014vxa,Harigaya:2014fca}.

We also show in this paper that our 2-parameter SUGRA model which we relate to the 2-parameter 
$\frac{1}{M^{2}} R^{\beta}$ model is the most economical representation of the 5-parameter scalar-curvature
coupled inflation models $\xi \phi^{a} R^{b} + \lambda \phi^{4(1+\gamma)}$ in terms of the number of parameters.

The organization of this paper is as follows: In the Section (\ref{power-law}), we calculate an equivalent 
scalar potential in the Einstein frame for $R+ \frac{1}{M^{2}} R^{\beta}$ gravity. We then find the parameter $M$ and $\beta$
values which satisfy the observed amplitude $\Delta_{\cal R}^{2}$, spectral index $n_s$ and tensor to scalar
$r$. We fix model parameters for two cases: one with running of $n_{s}$ and another without running of $n_{s}$.
In the Section (\ref{SUGRA}), we give a SUGRA embedding of the
$\frac{1}{M^{2}}R^{\beta}$ model with a specific choice of the K\"ahler potential $K$ and superpotential $W$. In the Section 
(\ref{higgs-inflation}), we show that the generalized curvature coupling model $\xi \phi^{a} R^{b} + \lambda \phi^{4(1+\gamma)}$ 
is equivalent to $ R+ \frac{1}{M^{2}} R^{\beta}$ model and give the relation between the parameters of these two
generalized models. Finally we conclude in Section (\ref{concl}).
\section{ Power-law Starobinsky model}\label{power-law}

We start with a $f(R)$ action of the form \cite{DeFelice:2010aj,Nojiri:2010wj}
\begin{equation}
S_{J} = \frac{-M_p^{2}}{2}\int d^4x\sqrt{-g} \left(R+ \frac{1}{6M^{2}} \frac{R^{\beta}}{M_p^{2\beta-2}} \right)
\label{eq1}
\end{equation}
where $M_p^2 = (8\pi G)^{-1}$, $g$ is the determinant of the metric $g_{\mu\nu}$ and $M$ is a
dimensionless real parameter. The subscript $J$ refers to Jordan frame which indicates that the gravity sector 
is not the Einstein gravity form. The action~(\ref{eq1}) can be transformed to an Einstein frame
action using the conformal transformation $\tilde{g}_{\mu\nu}(x) = \Omega(x) g_{\mu\nu}(x)$, where
$\Omega$ is the conformal factor and tilde represents quantities in the Einstein frame. Under conformal 
transformation the Ricci scalar $R$ in the two frames is related by
\ba
R=\Omega\big(\tilde{R}+3\tilde{\square} \omega-\frac{3}{2}\tilde{g}^{\mu\nu}
\partial_{\mu} \omega \partial_{\nu} \omega \big) \label{Rex}
\ea
where $\omega \equiv \ln \Omega$.
\begin{figure}[t!]
 \centering
\includegraphics[width= 7.0cm]{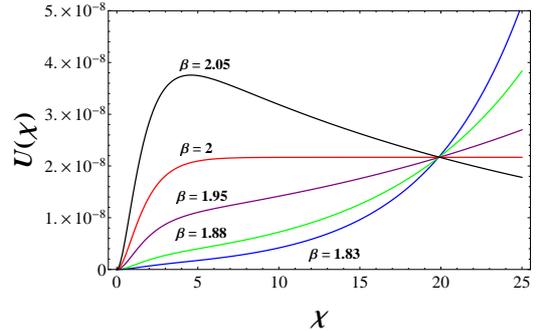}\hspace{0.5cm}
\caption
   {The nature of the potential (\ref{U1}) for different $\beta$ values (with $M=1.7\times10^{-4}$). 
   The potential and the field values are in $M_p=1$ units.}
\label{fig1}
\end{figure}
If one choose the conformal factor to be $\Omega = F = \frac{\partial f(R)}{\partial R}$ and introduce a new scalar field $\chi$ defined by 
$\Omega \equiv \exp\left(\frac{2\chi}{\sqrt{6} M_p}\right)$, using (\ref{Rex}), the action (\ref{eq1}) gets transform to an Einstein Hilbert form:
\begin{equation}
S_{E}=\int d^{4}x\sqrt{-\tilde{g}}\left[\frac{-M_p^{2}}{2}\tilde{R}
+\frac{1}{2}\tilde{g}^{\mu\nu}\partial_{\mu}\chi\partial_{\nu}\chi+U(\chi)\right]
\label{Ein}
\end{equation}
where $U(\chi)$ is the Einstein frame potential given by
\begin{equation}
U(\chi)=\frac{\left(R F(R)-f(R)\right)M_p^{2}}{2 F(R)^2}\,
\label{potentialein}
\end{equation}
which, by using the $f(R)$ form (\ref{eq1}) and $\Omega=F=\exp\left(\frac{2\chi}{\sqrt{6} M_p}\right)$, can be
given explicitly in terms of model parameters $M$ and $\beta$ as
\ba
U(\chi) &=& \frac{(\beta-1)}{2} \left(\frac{6M^{2}}{\beta^{\beta}}\right)^{\frac{1}{\beta-1}} 
\exp\left[ \frac{2\chi}{\sqrt6}\bigg(\frac{2-\beta}{\beta-1}\bigg)\right]
\nonumber\\
&&\times\left[1 - \exp\bigg(\frac{-2\chi}{\sqrt6}\bigg)\right]^{\frac{\beta}{\beta-1}}\label{U1}
\ea
where we have taken $M_p=1$ and from here onwards we shall work in $M_p=1$ units.
Also we see that in the limit $\beta\rightarrow 2$ potential (\ref{U1}) reduces to exponentially 
corrected flat plateau potential of the Starobinsky model.

Assuming large field limit $\chi \gg \frac{\sqrt6}{2}$ and $ 1 < \beta < 2$, the potential (\ref{U1}) reduces to
\ba
U(\chi) \simeq \frac{(\beta-1)}{2} \left(\frac{6M^{2}}{\beta^{\beta}}\right)^{\frac{1}{\beta-1}} \exp\Bigg[\frac{2\chi}{\sqrt6}
\bigg(\frac{2-\beta}{\beta-1}\bigg)\Bigg]\label{U2}
\ea
We shall use eq.(\ref{U2}) later in the Section (\ref{SUGRA}) to compare with SUGRA version of the power law potential in the large field limit.

In Fig.\ref{fig1} we plot the potential for small deviations from the Starobinsky model value $\beta=2$. 
We see that the potential is very flattest  for $\beta=2$ but becomes very steep even with small deviation 
from Starobinsky model value $\beta=2$. The scalar curvature perturbation $\Delta^{2}_{\mathcal{R}} \propto \frac{U(\chi)}{\epsilon}$ is
fixed from observations which implies that the magnitude of the potential $U(\chi)$ would have to be larger as 
$\epsilon$ increases for steep potential. The tensor perturbation which depends on the magnitude of
$U(\chi)$ therefore increases rapidly as $\beta$ varies from $2$. The variation of $r$ 
with $\beta$ is shown in the Fig.\ref{fig3}.

From eq.(\ref{U1}), in the large field approximation, the slow roll parameters in Einstein frame
can be obtained as
\ba
\epsilon &=& \frac{1}{2}\left(\frac{U'}{U}\right)^2 \nonumber\\
&\simeq& \frac{1}{3}\left[\frac{\beta(3-2\beta)}{(\beta-1)^{2}} \exp\left(\frac{-2\chi}{\sqrt{6}}\right) + \frac{\beta-2}{\beta-1}\right]^{2} \label{slow1}
\ea
\ba
\eta &=& \frac{U''}{U} \nonumber\\
&\simeq& \frac{-2}{3} \left[\frac{\beta(3-2\beta)^{2}}{(\beta-1)^{3}} \exp\left(\frac{-2\chi}{\sqrt{6}}\right) - \frac{(\beta-2)^{2}}{(\beta-1)^{2}}\right]
\ea
\ba
\xi &=& \frac{U'U'''}{U^{2}} \nonumber\\
&\simeq& \frac{4\sqrt{\epsilon}}{3\sqrt{3}} \left[\frac{\beta(3-2\beta)^{3}}{(\beta-1)^{4}} \exp\left(\frac{-2\chi}{\sqrt{6}}\right) + \frac{(\beta-2)^{3}}{(\beta-1)^{3}} \right]
\ea
%
%
The field value $\chi_{e}$ at the end of inflation can be fixed from eq.(\ref{slow1}) by using the end
of inflation condition $\epsilon \simeq 1$. And the initial scalar field value $\chi_{s}$ corresponding to
$N = 60$ e-folds before the end of inflation, when observable CMB modes leave the horizon,
can be fixed by using the e-folding expression $N=\int_{\chi_e}^{\chi_s} \frac{U(\chi)}{U'(\chi)} d\chi$.

Under slow-roll approximation we use the standard Einstein frame
relations for the amplitude of the curvature perturbation
$\Delta_{\mathcal R}^{2} = \frac{1}{24 \pi^{2}} \frac{U^{\ast}}{\epsilon^{\ast}}$, 
the spectral index $n_{s} = 1-6\epsilon^{\ast}+2\eta^{\ast}$, the running of spectral index
$\alpha_{s} = \frac{dn_{s}}{d\ln k} = 16\epsilon^{\ast}\eta^{\ast} -24(\epsilon^{\ast})^{2} - 2\xi^{\ast}$
and the tensor to scalar ratio $r = 16\epsilon^{\ast}$ to fix the parameters of our model. Note that the superscript $\ast$ 
indicates that the the observables are evaluated at the initial field value $\chi_{s}$.
%
 \begin{figure}[t!]
 \centering
\includegraphics[width= 7.0cm]{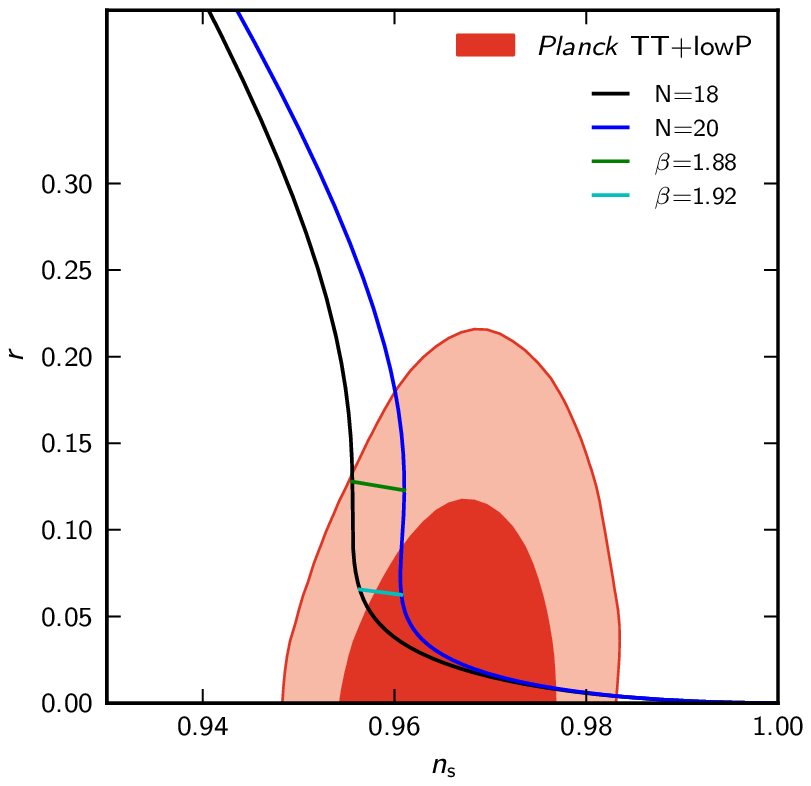}\hspace{0.5cm}
 \includegraphics[width= 7.0cm]{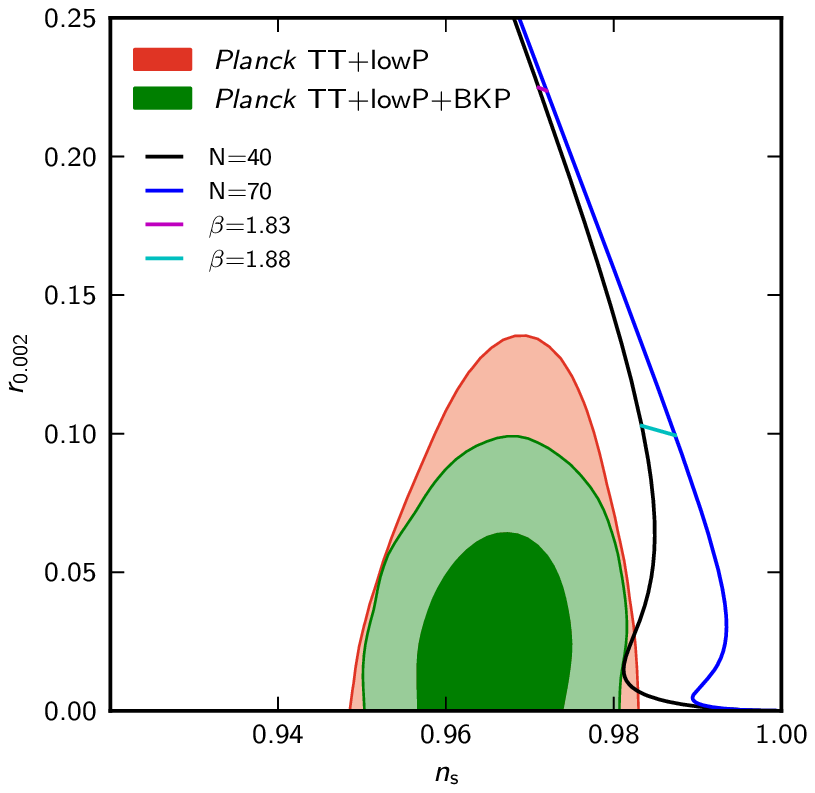}\hspace{0.0cm}
\caption
   {The regions of $(n_{s},r)$ allowed by Planck-2015 and joint BKP analysis at $68 \% CL$ and $95 \% CL$ are shown.
   In the upper panel running of $n_{s}$ is considered and in the lower panel there is no running of $n_{s}$.
   The colored contour lines are the predictions for our model for two sets of $\beta$ and $N$ values corresponding
   to $M\approx 10^{-4}$ which satisfies the observed amplitude of the CMB power spectrum.}
 \label{fig2}
\end{figure}

We know from CMB observations, for 8-parameter $\Lambda CDM$+$r$+$\alpha_{s}$ model, that if there is a large running of the
spectral index $\alpha_{s}= -0.013^{+0.010}_{-0.009}$ at ($68 \% CL$, PlanckTT+lowP) then the amplitude is
$10^{10}\ln(\Delta_{\mathcal R}^{2}) = 3.089\pm 0.072$, the spectral index is $n_{s}= 0.9667
\pm 0.0132$ and tensor to scalar ratio is $r_{0.05} < 0.168$ ($95 \% CL$, PlanckTT+lowP) \cite{BKP:2015,Ade:2015lrj,Planck:2015}.
Also a joint BICEP2/Keck Array and Planck analysis put an upper limit on $r_{0.05}<0.12 (95\% CL)$.
Since the scalar potential $U(\chi)$ depends on both the parameters $M$ and $\beta$ whereas the slow roll
parameters depend only on $\beta$, therefore parameter $M$ affects only the scalar amplitude 
$\Delta_{\cal R}^{2} \propto \frac{U(\chi)}{\epsilon}$ whereas $r$, $n_{s}$ and $\alpha_{s}$ which depend
only on slow roll parameters remain unaffected by $M$. Therefore taking amplitude from the observation and
fixing the number of e-foldings $N$ fixes the value of $M$ and $\beta$.
We find numerically that for the best fit parameter values $\beta \simeq 1.88$ and $M \simeq 1.7 \times 10^{-4}$, 
the e-foldings turns out to be $N \approx 20$. The tensor to scalar ratio can be
further reduced to $r\approx 0.06$ for $\beta \simeq 1.92$, $M \simeq 10^{-4}$ but e-foldings still comes out to be low
$N\approx 20$, see Fig.\ref{fig2}(upper panel). Therefore constraining model parameters using running data implies
that cosmological problems like Horizon and flatness problems which require a minimum of $50-60$ e-foldings cannot
be solved with the power law generalization of the Starobinsky model.

Also from CMB observations, for 7-parameter $\Lambda CDM$+$r$ model, when there is no scale dependence 
of the scalar and tensor spectral indices the bound on $r$ becomes tighter  $r_{0.002} < 0.1$ ($95 \% CL$, PlanckTT+lowP) and
the amplitude and the spectral index become $10^{10}\ln(\Delta_{\mathcal R}^{2}) = 3.089\pm 0.036$ and $n_{s}= 0.9666
\pm 0.0062$ respectively at ($68 \% CL$, PlanckTT+lowP) \cite{BKP:2015,Ade:2015lrj,Planck:2015}.
We find that the values of $M \simeq 1.7\times10^{-4}$ and $\beta \simeq 1.83$ which satisfy 
the amplitude and the spectral index for $N\approx 60$ gives large $r\approx 0.22$. Also we see that for $\beta \simeq 1.88$
and $M\simeq 1.25\times 10^{-4}$ tensor to scalar ratio can be reduced to $r\simeq 0.1$ but it increases $n_{s}\simeq 0.985$, see Fig.\ref{fig2}(lower panel).
\begin{figure}[t!]
 \centering
\includegraphics[width= 7.0cm]{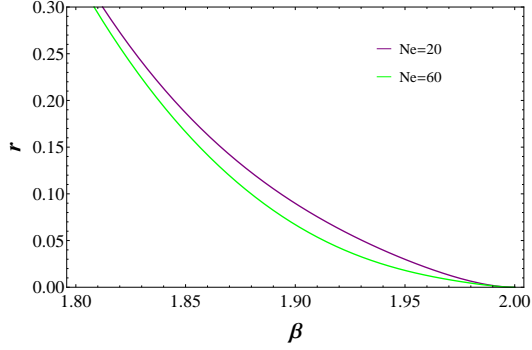}\hspace{0.0cm}
\caption
   {The variation of $r$ with $\beta$ shown for two cases studied in our model: (i) for $N=20$ when running of $n_{s}$ 
   is considered and (ii) for $N=60$ when there is no running of $n_{s}$.}
\label{fig3}
\end{figure}
\section{Power law Starobinsky model from supergravity}\label{SUGRA}
 
In this section we give a SUGRA model of the power law Starobinsky model. We shall derive a model
where the scalar potential in the Einstein frame is the same as eq.(\ref{U2}) which we have shown
in the Section (\ref{power-law}) is equivalent to the power law Starobinsky model 
$R+ \frac{1}{6M^{2}} R^{\beta}$. The F-term scalar potential in SUGRA depends upon the 
combination of the K\"ahler potential $K(\Phi_{i})$ and the superpotential 
$W(\Phi_{i})$ as $G \equiv K + \ln W +\ln W^{\ast}$, where $\Phi_{i}$ are the chiral superfields
whose scalar component are $\phi_{i}$ \cite{Cremmer:1978hn}.
The effective potential and kinetic term in the Einstein frame are given by
\ba
V=e^{G}\left[\frac{\partial G}{\partial \phi^{i}} K^{i}_{j*} \frac{\partial G}{\partial \phi^{*}_{j}} - 3 \right] \label{LV}
\ea
and
\ba
\mathcal{L}_{K}=K_{i}^{j*} \partial_{\mu}\phi^{i} \partial^{\mu}\phi^{*}_{j} \label{LK}
\ea
respectively, where $K^{i}_{j*}$ is the inverse of the K\"ahler metric $K_{i}^{j*} \equiv \partial^{2}K / \partial\phi^{i}\partial\phi^{*}_{j}$.

A no-scale SUGRA model \cite{Ellis:2013xoa} with a choice of the K\"ahler potential $K = -3 \ln \left[T+T^{*}-\phi\phi^{*}/3\right]$ and
a minimal Wess-Zumino superpotential with a single chiral superfield $\Phi$
\ba
W(\Phi)= \frac{\mu}{2} \Phi^{2}- \frac{\lambda}{3} \Phi^{3} \label{SP}
\ea
gives the same F-term  potential in the Einstein frame as the  Starobinsky model which 
give vanishing tensor to scalar ratio $r\sim 0.003$ for specific choice $\frac{\lambda}{\mu}=\frac{1}{3}$. A slight change in the
ratio $\frac{\lambda}{\mu}$ can increase $r$ upto $r\sim 0.005$ but it gives large $n_{s}\approx 0.98$.

To get a no-scale SUGRA model corresponding to power law Starobinsky model which can give a larger $r$,
we choose the minimal Wess-Zumino form of the superpotential (\ref{SP})
and a minimal no-scale K\"ahler potential with an added $(\phi+\phi^{*})^{n}$ term as
\ba
K = -3 \ln \left[T+T^{*}-\frac{(\phi+\phi^{*})^{n}}{12}\right] \label{KP}
\ea
which can be motivated by a shift symmetry $T\rightarrow T + i C$, $ \phi \rightarrow \phi +i C $ with $ C $ real,
on the K\"ahler potential. Here $T$ is a modulus field and $\phi$ is a matter filed which plays the role of inflaton.
%

We calculate eq.(\ref{LV}) and eq.(\ref{LK}) for chosen K\"ahler potential  (\ref{KP}) and superpotential (\ref{SP}).
We assume that the $T$ field gets a vev $\langle T + T^{*} \rangle = 2 \langle Re T \rangle = c >0$ and 
 $\langle ImT \rangle =0 $. We write $\phi$ in terms of its real and imaginary parts $\phi=\phi_{1}+i\phi_{2}$.
If we fix the imaginary part of the inflaton field $\phi$ to be zero then $\phi=\phi^{*}=\phi_{1}$ and for simplicity we
replace $\phi_{1}$ by $\phi$, the effective Lagrangian in the Einstein frame is given by
 \ba
{\cal{L}}_{E} &=& \frac{n (2\phi)^{n-2} [c(n-1) + \frac{(2\phi)^{n}}{12}]}{4[c - \frac{(2\phi)^{n}}{12}]^{2}}
\left|\partial_{\mu}\phi \right|^{2} \nonumber\\
&&- \frac{4(2\phi)^{2-n}}{n(n-1)[c-\frac{(2\phi)^{n}}{12}]^{2}}
\left|\frac{\partial{W}}{\partial{\phi}}\right|^{2} \label{LE1}
\ea
To make the kinetic term canonical in the ${\cal{L}}_{E}$, we redefine the field $\phi$ to $\chi$ with
\ba
\frac{\partial \chi}{\partial \phi} = -\frac{\sqrt{n(2\phi)^{n-2}[c(n-1)+\frac{(2\phi)^{n}}{12}]}}{2[c - \frac{(2\phi)^{n}}{12}]} \label{chiphi}
\ea
Assuming that $n \sim \mathcal{O}(1)$ and the large field limit $ (2\phi)^{n} \gg 12c $ during inflation, integrating eq.(\ref{chiphi}) gives 
\ba
\phi \simeq \frac{1}{2} \exp\left(\frac{2\chi}{\sqrt{3n}}\right) \left[1 + \frac{6c(n+1)}{n}
\exp\left(\frac{-2 n\chi}{\sqrt{3n}}\right) \right] \label{phi1}
\ea
Now substituting from eq.(\ref{SP}) and eq.(\ref{phi1}) into the potential term of eq.(\ref{LE1}) and simplifing, we get 
the effective scalar potential in the Einstein frame as
\ba
V &=& \frac{144\mu^{2}}{n(n-1)} \exp\left[\frac{2\chi}{\sqrt{6}}\left(\frac{3\sqrt{2}(2-n)}{\sqrt{n}}\right)\right]
\nonumber\\
&&\times \left[1- \frac{2\mu}{\lambda}\exp\left(\frac{-2\chi}{\sqrt{3n}}\right) -\frac{9c(n^{2}-n-2)}{n} \right. \nonumber\\
&& \left. \times \exp\left(\frac{-2n\chi}{\sqrt{3n}}\right) \right]^{2} \label{V3}
\ea
which, assuming $1<n<2$, in the large field limit $\chi \gg \frac{\sqrt{3n}}{2}$ is equivalent to
\ba
V \simeq \frac{144\mu^{2}}{n(n-1)} \exp\left[\frac{2\chi}{\sqrt{6}}\left(\frac{3\sqrt{2}(2-n)}{\sqrt{n}}\right)\right] \label{V4}
\ea
We see that in the limit $n \rightarrow 2$ and with the specific choice  $\frac{\lambda}{\mu}=\frac{1}{2}$,
the potential (\ref{V3}) reduces to Starobinsky Model potential.

We can now compare the power law potential (\ref{U2}) and SUGRA potential (\ref{V4}) for inflaton to show the relation
between the parameters of the two model. Comparing the constant coefficient and exponent in the two potentials we get
\ba
\beta = \frac{2\sqrt{n}+3\sqrt{2}(2-n)}{\sqrt{n}+3\sqrt{2}(2-n)}\label{beta1}
\ea
and
\ba
M^{2} = \frac{\beta^{\beta}}{6} \left[\frac{288 \mu^{2}}{n(n-1)(\beta-1)}\right]^{\beta-1}. \label{M1}
\ea
Numerically we find the SUGRA model parameter values (in $M_p = 1$ unit) for three values of $\beta$ corresponding to
running and without running of spectral index $n_{s}$ as depicted in Fig.\ref{fig2} and for Starobinsky limit $\beta=2$
as shown in the TABLE[\ref{sugrapara}].
\begin{table}
\small
\centering
    \begin{tabular}{|l|l|l|l|l|}
        \hline
        $ \beta $ & $ \ M$ & $ n  $ &  $ \mu=\frac{|\lambda|}{2} $ & $ \alpha_{s}= \frac{dn_{s}}{d\ln{k}}$ \\ \hline
        1.83 & $ 1.7\times 10^{-4} $ & $ 1.93 $ &  $ 3.13\times 10^{-6} $ & $ -9.16\times 10^{-6} $ \\ \hline
        1.88 & $ 1.7\times 10^{-4} $  & $ 1.96  $ & $ 5.54 \times 10^{-6} $ & $ -2.86\times 10^{-3} $\\ \hline
        2.00 & $ 1.1 \times 10^{-5}  $ & $ 2.00  $ &   $ 1.16 \times 10^{-6} $ & $ -5.23\times 10^{-4} $\\ \hline
   \end{tabular}
    \caption{The SUGRA model parameter values (in $M_p = 1$ unit) for three values of $\beta$ corresponding to
    running and without running of spectral index $n_{s}$ as depicted in Fig.\ref{fig2} and for Starobinsky limit $\beta=2$.}
    \label{sugrapara}
\end{table}
\section{Equivalence of the Power-law Starobinsky Model with generalized non-minimally curvature coupled  models }\label{higgs-inflation}

In this section we will show that generalized non-minimally coupled Inflation models $\xi \Phi^a R^b$ \cite{girish}
with the quantum corrected $\Phi^{4}$-potential \cite{Joergensen:2014rya,Codello:2014sua,Gao:2014fha}
can be reduced to the power law Starobinsky form. We consider the generalised non-minimal coupling $\xi \Phi^a R^b$ and the quantum correction to 
quartic scalar potential $\Phi^{4(1+\gamma)}$ into the action
\ba
S_{J} &=& \int d^4x \sqrt{-g} \left(-\frac{M_p^2 R}{2}-\frac{\xi \Phi^{a}R^{b}}{2 M_p^{a+2b-4}}\right. \nonumber\\
&& \left. + \frac{1}{2}g^{\mu\nu}\partial_{\mu}\Phi\partial_{\nu}\Phi +
\frac{\lambda \Phi^{4(1+\gamma)}}{4 M_{p}^{4\gamma}}\right)\label{action_J1}
\ea
where the scalar field $\Phi$ is the inflaton field.
Since during inflation potential energy of the scalar field is dominant therefore kinetic term in the action
$S_{J}$ can be neglected w.r.t. potential, the action reduces to
\ba
\int d^4x \sqrt{-g} \left(- \frac{M_p^2 R}{2}-\frac{\xi \Phi^{a}R^{b}}{2 M_p^{a+2b-4}}+\frac{\lambda \Phi^{4(1+\gamma)}}{4 M_{p}^{4\gamma}}\right)\label{action_J2}
\ea
we may integrate out the scalar field through its equation of motion  $\frac{\partial L}{\partial \Phi} \approx 0$ \cite{stelle}, which implies 
\ba
\Phi \approx \left(\frac{\xi a R^{b}}{2\lambda (1+\gamma) M_p^{a+2b-4(1+\gamma)}}\right)^{\frac{1}{4(1+\gamma)-a}} \label{Phi1}
\ea
Using eq.(\ref{Phi1}) for $\Phi$, the action (\ref{action_J2}) reduces to power law Starobinsky action
\ba
\int d^4x \sqrt{-g} \left(\frac{-M_p^{2}}{2}\right) \left( R + \frac{1}{6M^{2}} \frac{R^{\beta}}{M_p^{2\beta-2}} \right)
\ea
 where the two parameters $\beta$ and $M$ of the power law model are identified in terms of $a$, $b$, $\lambda$, $\xi$ and $\gamma$ as
 \ba
  \beta=\frac{4b(1+\gamma)}{4(1+\gamma)-a}\label{beta2}
 \ea 
 and
 \ba
 M^{2} = \frac{a}{3(4(1+\gamma)-a)\lambda}\left(\frac{2\lambda (1+\gamma)}{\xi a} \right)^{\frac{4(1+\gamma)}{4(1+\gamma)-a}} \label{M2}
 \ea  
 which for $a=2,~ b=1, \gamma=0$, i.e at $\beta = 2$, reduces to Higgs Inflation-Starobinsky case
 $M_S^{2}\approx \frac{\lambda}{3\xi^{2}} \approx 10^{-10}$.
 Also with $a=2, b=1, \gamma \neq 0$ results of the references \cite{Joergensen:2014rya,Codello:2014sua} are obtained.
\section{Conclusion} \label{concl}

We have explored a generalization of the Starobinsky model with a $\frac{1}{M^{2}}R^{\beta}$
model and fit $\beta $ and $M$ from CMB data. We find that to fit the amplitude
$\Delta_{\mathcal R}^{2}$ and the spectral index $n_{s}$  (with no running) from observations
\cite{BKP:2015,Ade:2015lrj,Planck:2015} we require $M \simeq 1.7 \times 10^{-4}$ 
and $\beta \simeq 1.83$ for $N \approx60$ but these parameter values gives large $r\approx 0.22$.
Also we find that the parameters $\beta$ and $M$ deviates from the
$M \approx 10^{-5}$ and $\beta = 2$ of the original Starobinsky model which could fit the amplitude 
and the spectral index but predicted very small value of $r \sim 10^{-3}$. When large running
of the spectral index $\alpha_{s}\sim 10^{-3}$ is considered we find that the best fit parameter
values are $\beta \simeq 1.88$ and $M \simeq 1.7 \times 10^{-4}$ which
gives $N \approx 20$. This implies that the standard cosmological problems like Horizon and flatness problems which 
require a minimum of $50-60$ e-foldings cannot be solved with the power law generalization
of the Starobinsky model.

We have shown that the 5-parametes generalised non-minimal scalar-curvature coupled inflation models
with the quantum correction to quartic scalar potential i.e. $\xi \Phi^a R^b + \Phi^{4(1+\gamma)}$
are actually equivalent to 2-parameter power law Starobinsky model $\frac{1}{M^{2}}R^{\beta}$.
Therefore we see that in terms of number of parameters the power law model is the most economical 
parametrization of the class of scalar-curvature models with quantum corrected $\Phi^{4}$-potential.

In this paper we have given a SUGRA model for the general power law $\frac{1}{M^{2}}R^\beta$ model by adding a 
$(\Phi+\bar {\Phi})^n$ term to the minimal no-scale K\"ahler potential and with a Wess-Zumino
form of the superpotential $W(\Phi)$. In the limit $n=2$ the Starobinsky limit $\beta=2$ 
is obtained. We derive the relations between the two parameters of the power-law Starobinsky
model and the two parameters of our SUGRA model. The interesting point about the generalisation
is that the small deviations from the Starobinsky limit of $n=\beta=2$ can produce value of
$r\sim 0.1$ which is consistent with the joint Planck+BICEP2+Keck Array upper bound on $r<0.12 (95\%CL)$.
Generalisations of the Starobinsky model which can explain a possible larger value of $r$
are therefore of interest.
\begin{center}
 {\bf Acknowledgement}
 \end{center}
We thank Akhilesh Nautiyal for valuable discussions.

\end{document}